\pdfoutput=1
\documentclass[11pt,twoside,a4paper,cmspaper,final,collab]{cms-tdr}
\def\svnVersion{5dd70a6}\def\svnDate{2025/01/10}\def\cmsCernNoTag{CERN-EP-2024-145}\def\cmsCernDate{\today}\def\cmsMessage{Published in Physics Letters B as \href{http://dx.doi.org/10.1016/j.physletb.2024.139210}{\doi{10.1016/j.physletb.2024.139210}.}}
\begin{document}\cmsNoteHeader{HIG-23-006}

\newcommand{\HH}{\ensuremath{{\PH\PH}}\xspace}
\newcommand{\ggF}{\ensuremath{\Pg\Pg\mathrm{F}}\xspace}
\newcommand{\VBF}{\ensuremath{\mathrm{VBF}}\xspace}
\newcommand{\ggHH}{\ensuremath{{\ggF~\PH\PH}}\xspace}
\newcommand{\qqHH}{\ensuremath{{\VBF~\PH\PH}}\xspace}
\newcommand{\VBFHH}{\qqHH}
\newcommand{\VHH}{\ensuremath{\PV\PH\PH}\xspace}

\newcommand{\bbgg}{\ensuremath{\gamma\gamma\bbbar}\xspace}
\newcommand{\bb}{\ensuremath{\bbbar}\xspace}
\newcommand{\bbbb}{\ensuremath{\bbbar\bbbar}\xspace}
\newcommand{\bbtt}{\ensuremath{\tau\tau\bbbar}\xspace}
\newcommand{\bbZZ}{\ensuremath{\PZ\PZ\bbbar}\xspace}
\newcommand{\bbWW}{\ensuremath{\PW\PW\bbbar}\xspace}
\newcommand{\WWgg}{\ensuremath{\PW\PW\gamma\gamma}\xspace}
\newcommand{\HHbbgg}{\ensuremath{\PH\PH\to\gamma\gamma\bbbar}\xspace}
\newcommand{\HHbbtt}{\ensuremath{\PH\PH\to\tau\tau\bbbar}\xspace}
\newcommand{\HHbbWW}{\ensuremath{\PH\PH\to\PW\PW\bbbar}\xspace}
\newcommand{\HHbbZZllll}{\ensuremath{\PH\PH\to\bbbar\PZ\PZ(4\Pl)}\xspace}
\newcommand{\HHbbbb}{\ensuremath{\PH\PH\to \bbbb}\xspace}
\newcommand{\HHmultilep}{\ensuremath{\PH\PH \to \mathrm{leptons}}\xspace}
\newcommand{\ppHHbbgg}{\ensuremath{\Pp\Pp\to\PH\PH\to\gamma\gamma\bbbar}\xspace}
\newcommand{\HVV}{\ensuremath{\PH\to\PV\PV}\xspace}
\newcommand{\VHHbb}{\ensuremath{\PH\PH \to \bbbb~(\PV\PH\PH)}\xspace}
\newcommand{\ggHHbbbb}{\ensuremath{\PH\PH \to \bbbb~(\ggF)}\xspace}
\newcommand{\qqHHbbbb}{\ensuremath{\PH\PH \to \bbbb~(\VBF)}\xspace}

\newcommand{\SH}{\ensuremath{\mbox{single-}{\PH}}\xspace}
\newcommand{\singleH}{\SH}
\newcommand{\ggH}{\ensuremath{\ggF~\PH}\xspace}
\newcommand{\VBFH}{\ensuremath{{\VBF~\PH}}\xspace}
\newcommand{\qqH}{\VBFH}
\newcommand{\VH}{\ensuremath{\PV\PH}\xspace}
\newcommand{\ZH}{\ensuremath{\PZ\PH}\xspace}
\newcommand{\WH}{\ensuremath{\PW\PH}\xspace}
\newcommand{\ttH}{\ensuremath{\ttbar\PH}\xspace}
\newcommand{\tHq}{\ensuremath{\PQt\PH\PQq}\xspace}
\newcommand{\tH}{\ensuremath{\PQt\PH}\xspace}

\newcommand{\Hgg}{\ensuremath{\PH\to\gamma\gamma}\xspace}
\newcommand{\Hbb}{\ensuremath{\PH\to \bbbar}\xspace}
\newcommand{\Hcc}{\ensuremath{\PH\to \ccbar}\xspace}
\newcommand{\HZZllll}{\ensuremath{\PH\to 4\Pl}\xspace}
\newcommand{\HZg}{\ensuremath{\PH\to \PZ\gamma}\xspace}
\newcommand{\HWW}{\ensuremath{\PH\to \PW \PW}\xspace}
\newcommand{\Hmm}{\ensuremath{\PH\to \PGm \PGm}\xspace}
\newcommand{\Htt}{\ensuremath{\PH\to \PGt \PGt}\xspace}
\newcommand{\Hinv}{\ensuremath{\PH\to \text{invisible}}\xspace}
\newcommand{\ttHgg}{\ensuremath{\PH \to \PGg\PGg~(\ttH)}\xspace}
\newcommand{\ttHbb}{\ensuremath{\PH \to \bbbar~(\ttH)}\xspace}
\newcommand{\ttHmultilep}{\ensuremath{\PH \to \text{leptons}~(\ttH)}\xspace}
\newcommand{\VHgg}{\ensuremath{\VH(\PGg\PGg)}\xspace}
\newcommand{\VHbb}{\ensuremath{\PH \to \bbbar~(\VH)}\xspace}
\newcommand{\qqHgg}{\ensuremath{\qqH(\PGg\PGg)}\xspace}
\newcommand{\ggHgg}{\ensuremath{\ggH(\PGg\PGg)}\xspace}
\newcommand{\ggHbb}{\ensuremath{\PH \to \bbbar~(\ggF)}\xspace}
\newcommand{\tHqgg}{\ensuremath{\PQt\PH(\PGg\PGg)\PQq}\xspace}

\newcommand{\eeee}{\ensuremath{\Pe\Pe\Pe\Pe }\xspace}
\newcommand{\eemm}{\ensuremath{\Pe\Pe\PGm \PGm }\xspace}
\newcommand{\mmmm}{\ensuremath{\PGm \PGm \PGm \PGm }\xspace}
\newcommand{\Zee}{\ensuremath{\PZ \to \Pe\Pe }\xspace}
\newcommand{\Zmumu}{\ensuremath{\PZ \to \PGm\PGm}\xspace}
\newcommand{\Zll}{\ensuremath{\PZ \to ll}\xspace}
\newcommand{\PHiggs}{\ensuremath{\PH}}
\newcommand{\muth}{\ensuremath{\PGt_\PGm\tauh}\xspace}
\newcommand{\elemu}{\ensuremath{\PGt_\Pe\PGt_\PGm}\xspace}
\newcommand{\eleth}{\ensuremath{\PGt_\Pe\PGt_{{\text h}}}\xspace}
\newcommand{\thth}{\ensuremath{\tauh\tauh}\xspace}

\newcommand{\mH}{\ensuremath{m_{\PH}}\xspace}
\newcommand{\mHH}{\ensuremath{m_{\PH\PH}}\xspace}
\newcommand{\mreg}{\ensuremath{m_\mathrm{reg}}\xspace}
\newcommand{\Dbb}{\ensuremath{D_{\bbbar}}\xspace}
\newcommand{\PTH}{\ensuremath{p_{\mathrm{T}}^{\PH}}\xspace}
\newcommand{\PTV}{\ensuremath{p_{\mathrm{T}}^{\PV}}\xspace}

\newcommand{\sigmaHH}{\ensuremath{\sigma_{\PH\PH}}\xspace}
\newcommand{\sigmaggHH}{\ensuremath{\sigma_{\ggHH}}\xspace}
\newcommand{\sigmaqqHH}{\ensuremath{\sigma_{\qqHH}}\xspace}
\newcommand{\sigmaVBFHH}{\sigmaqqHH}
\newcommand{\BR}{\ensuremath{\mathcal{B}(\HH\to\bbgg)}}
\newcommand{\lambdahhh}{\ensuremath{\lambda_3}\xspace}
\newcommand{\lambdahhhSM}{\ensuremath{\lambda_3^\mathrm{SM}}\xspace}
\newcommand{\lambdahhhhSM}{\ensuremath{\lambda_4^\mathrm{SM}}\xspace}
\newcommand{\lambdahhhh}{\ensuremath{\lambda_4}\xspace}
\newcommand{\kappalambda}{\ensuremath{\kappa_\lambda}\xspace}
\newcommand{\kl}{\kappalambda}
\newcommand{\kappat}{\ensuremath{\kappa_{\PQt}}\xspace}
\newcommand{\kb}{\ensuremath{\kappa_\PQb}\xspace}
\newcommand{\kc}{\ensuremath{\kappa_\PQc}\xspace}
\newcommand{\ktau}{\ensuremath{\kappa_\PGt}\xspace}
\newcommand{\kmu}{\ensuremath{\kappa_\PGm}\xspace}
\newcommand{\kF}{\ensuremath{\kappa_{f}}\xspace}
\newcommand{\kappav}{\ensuremath{\kappa_\PV}\xspace}
\newcommand{\kV}{\kappav}
\newcommand{\kW}{\ensuremath{\kappa_\PW}\xspace}
\newcommand{\kZ}{\ensuremath{\kappa_\PZ}\xspace}
\newcommand{\kVV}{\ensuremath{\kappa_{2\PV}}\xspace}
\newcommand{\ktwoV}{\ensuremath{\kVV}\xspace}
\newcommand{\HHVV}{\ensuremath{\PH\PH\PV\PV}\xspace}

\newcommand{\PGtHad}{\ensuremath{\PGt_\text{h}}\xspace}
\newcommand{\pp}{\ensuremath{\Pp{}\Pp}\xspace}

\ifthenelse{\boolean{cms@external}}{\providecommand{\cmsLeft}{upper\xspace}}{\providecommand{\cmsLeft}{left\xspace}} 
\ifthenelse{\boolean{cms@external}}{\providecommand{\cmsRight}{lower\xspace}}{\providecommand{\cmsRight}{right\xspace}} 

\cmsNoteHeader{HIG-23-006}
\title{Constraints on the Higgs boson self-coupling from the combination of single and double Higgs boson production in proton-proton collisions at \texorpdfstring{$\sqrt{s} = 13\TeV$}{sqrt(s) = 13 TeV}}

\date{\today}

\abstract{
The Higgs boson (\PH) trilinear self-coupling, \lambdahhh, is constrained via its  measured properties and limits on the HH pair production using the proton-proton collision data collected by the CMS experiment at $\sqrt{s} = 13\TeV$. The combination of event categories enriched in single-\PH and $\PH\PH$ events is used to measure \kl, defined as the value of \lambdahhh normalized to its standard model prediction, while simultaneously constraining the Higgs boson couplings to fermions and vector bosons. Values of \kl outside the interval $-1.2< \kl < 7.5$ are excluded at $2\sigma$ confidence level, which is compatible with the expected range of $-2.0< \kl < 7.7$ under the assumption that all  other Higgs boson couplings are equal to their standard model predicted values. Relaxing the assumption on the Higgs couplings to fermions and vector bosons  the observed (expected) \kl interval is constrained to be within $-1.4 < \kl < 7.8$ ($-2.3 < \kl < 7.8$) at $2\sigma$ confidence level.}
 
\hypersetup{
pdfauthor={CMS Collaboration},%
pdftitle={Constraints on the Higgs boson self-coupling from the combination of single and double Higgs boson production in proton-proton collisions at sqrt(s) = 13 TeV},%
pdfsubject={CMS},%
pdfkeywords={CMS, Higgs, HH, di-Higgs, self-coupling, combination}} 

\maketitle

\section{Introduction}
\label{sec:introduction}

\begin{figure}[ht!]
	\centering
	\includegraphics[width=.24\textwidth]{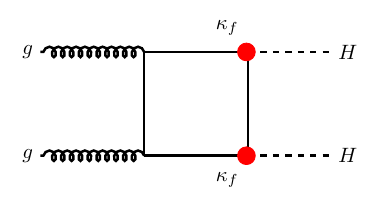}
	\includegraphics[width=.22\textwidth]{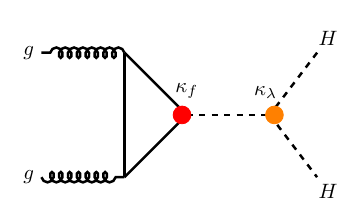}
	\caption{Feynman diagrams for the LO \HH production via gluon fusion. \kF corresponds to the modifier of the Higgs boson coupling strength (Section~\ref{sec:statistical_methods}) to fermions.}
	\label{fig:ggHHdiagrams}
\end{figure}

\begin{figure}[ht!]
	\centering
	\includegraphics[width=.2\textwidth]{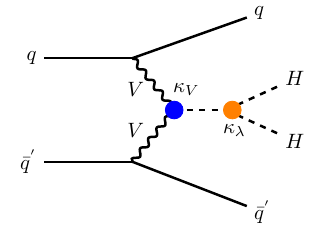}
	\quad
        \includegraphics[width=.2\textwidth]{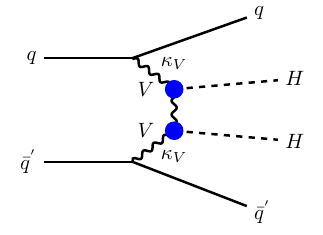}
	\quad
        \includegraphics[width=.2\textwidth]{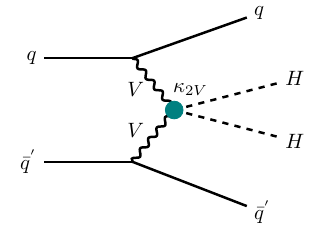}
	\caption{Feynman diagrams for the LO \HH production via vector boson fusion.  \kV and \ktwoV correspond to the modifiers of the Higgs boson coupling strength (Section~\ref{sec:statistical_methods}) to one and a pair of vector bosons, respectively.}
	\label{fig:qqHHdiagrams}
\end{figure}

Significant progress has been made in the study of the properties of the Higgs boson (H) since its discovery~\cite{PRLB2012-716-1-1,CMS-PAPERS-HIG-12-028,Chatrchyan:1529865,arXiv:2207.00043,arXiv:2207.00092}. The characterization of the Higgs boson is one of the primary goals of the CERN LHC physics program. This includes the measurement of its couplings to the vector bosons and fermions to test the compatibility with the standard model (SM) predictions. The Higgs boson plays a special role in several theories beyond the standard model (BSM), which could lead to deviations of the Higgs boson couplings from the SM predictions~\cite{Haber2015,BRANCO20121,Hespel2014,Binoth:1996au,PhysRevD.72.093007}. A crucial property of the Higgs boson predicted by the SM which is still only weakly constrained by the LHC data is its self-coupling. At low energy the potential $V$ of the Higgs boson field $H$ in the SM Lagrangian can be expressed as:
\begin{equation}\label{eq:LagrangianHiggsPot2}
V(H) = \frac{1}{2} \mH^2 H^2 + \lambdahhh v H^3 + \frac{\lambdahhhh}{4} H^4,
\end{equation}
where $\mH$ is the mass of the Higgs boson and $v \simeq 246$~GeV is the vacuum expectation value of the Higgs field. The form of the Higgs potential, hence the self-coupling constant, is tightly connected to the cosmological evolution of the Universe \cite{PhysRevD.110.030001}. The second and third term of Eq.~(\ref{eq:LagrangianHiggsPot2}) correspond to the trilinear and a quartic self-interactions of the Higgs boson. The parameters \lambdahhh and \lambdahhhh control the strength of the trilinear and quartic self-interactions of the Higgs boson, respectively. In the SM the strength of the quartic interaction \lambdahhhhSM is fully determined by the trilinear interaction \lambdahhhSM as: 
\begin{equation}
\lambdahhhhSM = \lambdahhhSM = \frac{\mH^2}{2v^2} \simeq 0.13.
\end{equation}
The modifier of the Higgs boson trilinear self-coupling with respect to the SM prediction is defined as $\kl = \lambdahhh / \lambdahhhSM$. The \kl value can be determined from the measurement of the Higgs boson pair production (\HH) cross section. In the SM, the main production mechanism for \HH in proton-proton collisions at $\sqrt{s} = 13\TeV$ is gluon fusion (\ggHH) with a cross section of $31.0^{+2.1}_{-7.2}$\unit{fb}~\cite{PhysRevD.58.115012,PhysRevLett.117.012001,Baglio.2018lrj,PhysRevLett.111.201801,Shao:2013bz,arXiv:1505.07122,Grazzini:2018bsd,PhysRevD.103.056002}. The second-largest \HH production mechanism is via vector boson fusion (\qqHH) with a cross section of $1.726 \pm 0.036$\unit{fb}~\cite{Baglio2013,FREDERIX2014142,PhysRevD.89.073001,PhysRevD.99.074028,PhysRevD.98.114016}. The Feynman diagrams corresponding to the \ggHH and \qqHH productions at the leading order (LO) are shown in Figs~\ref{fig:ggHHdiagrams} and \ref{fig:qqHHdiagrams}, respectively. The diagrams show that both mechanisms depend at the LO on \kl. A deviation of \kl from unity could significantly modify the cross sections and change the \HH kinematics. The most stringent observed (expected) upper limit on the \HH inclusive cross section previously set by the CMS Collaboration is 3.4 (2.5) times the SM prediction, and the observed constraint on \kl is $[-1.24, 6.49]$ at $95\% $ confidence level (\CL) assuming all the other Higgs boson couplings fixed to their SM value~\cite{arXiv:2207.00043}.

Another property of the Higgs boson predicted by the SM which can be probed through \HH searches is the coupling between two Higgs bosons and two massive vector bosons. The modifier of the \HHVV coupling strength (Section~\ref{sec:statistical_methods}), which is referred to as \ktwoV, affects the cross section of the \qqHH mechanism, and it is tightly related to the modifier of the strength of the coupling between one Higgs boson and a pair of massive vector bosons, \kV. In particular, the conservation of the asymptotic unitarity of the \qqHH  process imposes $\ktwoV = \kV^{2}$.
A deviation from this relationship would require the presence of BSM physics, e.g. a BSM particle, to maintain the asymptotic unitarity of the process or could significantly increase the rate of the \qqHH production. Allowing for such a deviation and assuming all the other Higgs boson couplings equal to their SM-predicted values, \ktwoV values smaller than or equal to zero are excluded at  a \CL corresponding to more than 5 standard deviations~\cite{arXiv:2207.00043}.

Other Higgs boson couplings also affect the \HH production cross sections. In particular at LO, the \HH cross section in the ggF production mode depends on the Higgs boson coupling to the top quark and the \qqHH cross section depends on the couplings of the Higgs boson to the vector bosons. Further dependencies on Higgs boson couplings arise when considering different \HH decay channels. The searches for \HH production generally provide weaker constraints on the Higgs boson couplings to fermions and vector bosons compared to \singleH measurements because of the much smaller cross sections.

\begin{figure}[ht!]
	\centering
	\includegraphics[width=.2\textwidth]{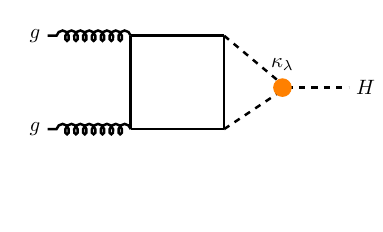}
	\quad
	\includegraphics[width=.2\textwidth]{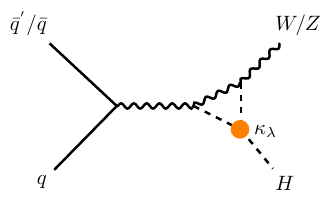}\\
	\ \\
	\includegraphics[width=.2\textwidth]{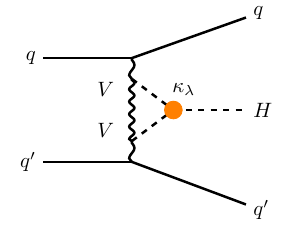}
	\quad
	\includegraphics[width=.2\textwidth]{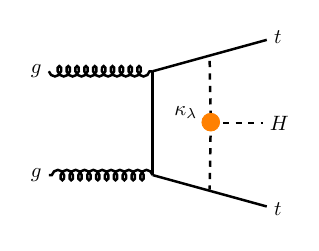}\\
   \ \\
	\includegraphics[width=.2\textwidth]{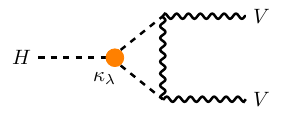}
	\quad
	\includegraphics[width=.2\textwidth]{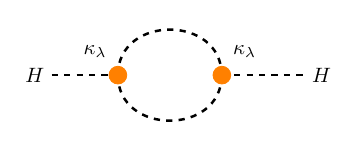}
	\caption{Feynman diagrams corresponding to \kl-dependent NLO corrections to the main \singleH production mechanisms (in the two top rows), to the \HVV decay width (bottom left) and to the Higgs boson propagator (bottom right).}
	\label{fig:singleHdiagrams}
\end{figure}

An alternative way to determine \kl is from the cross sections of the main \singleH production mechanisms utilizing next-to-LO (NLO) electroweak corrections that depend on \kl because of the appearance of loops and vertices with three Higgs bosons~\cite{Degrassi:2016wml}. These corrections affect the \singleH production cross sections and partial decay widths, as well as the Higgs boson propagator, as visible in the example Feynman diagrams of Fig.~\ref{fig:singleHdiagrams}.

The parametrizations of the Higgs boson production cross section and branching fraction variations as functions of \kl are provided in Refs~\cite{Degrassi:2016wml, arXiv:1709.08649v2}. The largest corrections affect the Higgs boson production in association with a \PZ or a \PW boson (\ZH and \WH, respectively, or generally \VH), a top quark-antiquark pair (\ttH), or a single top quark (\tH). In particular, the variation of the \ttH cross section is up to about $\pm10\% $ for \kl between $-$2 and 6. The variations of the Higgs boson production cross sections via gluon fusion (\ggH) and via vector boson fusion (\qqH) are milder but still relevant, i.e. up to about $-$3\%/+2\% for the same variation of \kl.
Additional information can be extracted from the \singleH differential cross sections. Some kinematic variables that are experimentally accessible, such as the Higgs boson transverse momentum (\PTH), are particularly sensitive to \kl. The \singleH production mechanisms that receive the largest \kl-dependent corrections to the \PTH spectrum are \ttH and \VH.

A framework to study the \singleH production cross sections with a well-defined phase space granularity is provided by the simplified template cross sections (STXS)~\cite{cernYR4,Delmastro:2669925}. The phase space divisions, known as \textit{stages}, provide varying levels of granularity, with the most detailed being \textit{stage} 1.2. Depending on the production mode, the measurements are performed in intervals of numbers of jets and \PTH, or for the \VH processes in intervals of the transverse momentum of the vector boson. The variations of the \singleH cross sections as functions of \kl with the granularity defined by the STXS of \textit{stage} 1.2 are provided in Ref.~\cite{LHCHWG-2022-002}.

The constraints on the Higgs boson couplings from the measurement of the \singleH cross sections are complementary to those from the \HH searches. In particular, the \singleH measurements provide stringent constraints on the Higgs boson couplings to fermions and vector bosons and a mild constraint on \kl, while the \HH searches provide tighter constraints on \kl.
Therefore, the inclusion of \singleH measurements improves the sensitivity with respect to \HH searches, where other Higgs boson couplings are assumed equal to their SM values, and it also enables more generic interpretations where constraints on \kl and other Higgs boson couplings are obtained from data simultaneously.

The combination of measurements and searches for the \singleH and \HH processes performed by the ATLAS Collaboration~\cite{ATLAS:2022jtk} excludes values of \kl outside the range $-0.4 < \kl < 6.3$ or $-1.4 < \kl < 6.1$ at $95\% $ \CL assuming the other Higgs boson couplings at their SM values or by treating the coupling modifiers \kappat, \kb, \ktau, and \kV as unconstrained parameters, respectively.

This letter presents a statistical combination of \singleH measurements and \HH searches to constrain the Higgs boson trilinear self-coupling under minimal assumptions on the Higgs boson couplings to fermions and to vector bosons. Constraints on \kl are also set assuming all the other Higgs boson couplings fixed to their SM values. The analyses considered are based on data from proton-proton (\pp) collisions produced by the LHC at $\sqrt{s} = 13\TeV$ and collected by the CMS experiment, corresponding to integrated luminosities of up to 138\fbinv~\cite{CMS-LUM-17-003, CMS-PAS-LUM-17-004,CMS-PAS-LUM-18-002}. The tabulated results are provided as HepData records~\cite{hepdata}.

The CMS apparatus~\cite{CMS:2008xjf,CMS:2023gfb} is a multipurpose, nearly hermetic detector, designed to trigger on~\cite{CMS:2020cmk,CMS:2016ngn} and identify electrons, muons, photons, and (charged and neutral) hadrons~\cite{CMS:2020uim,CMS:2018rym,CMS:2014pgm}. A global ``particle-flow'' algorithm~\cite{CMS:2017yfk} aims to reconstruct all individual particles in an event, combining information provided by the all-silicon inner tracker and by the crystal electromagnetic and brass-scintillator hadron calorimeters, operating inside a 3.8\unit{T} superconducting solenoid, with data from the gas-ionization muon detectors embedded in the flux-return yoke outside the solenoid. The reconstructed particles are used to build tau leptons, jets, and missing transverse momentum~\cite{CMS:2018jrd,CMS:2016lmd,CMS:2019ctu}.

\section{Analysis channels}
\label{sec:considered_analysis}

The  \singleH analyses considered are summarized in Table~\ref{tab:singleHanalyses}, which includes information on the production modes targeted by dedicated event categories and the maximum available phase space granularity of the cross section measurements, while the \HH searches included in the combination are listed in Table~\ref{tab:HHanalyses}. A more detailed description can be found in the references listed in Tables~\ref{tab:singleHanalyses} and \ref{tab:HHanalyses}. Most of the analyses considered are based on the \pp collision data collected between 2016 and 2018 with the CMS detector, corresponding to an integrated luminosity of 138~\fbinv, while two analyses use a subset of this data set.

The \Hcc searches \cite{Tumasyan:2842230,Tumasyan:2809290} are not included because major overlaps are expected with the considered \Hbb analysis~\cite{PhysRevLett.121.121801}. Searches sensitive to the Higgs couplings to the strange quark and first generation fermions \cite{CMS-PAS-HIG-23-005,Tumasyan:2823465} are not considered, since their contribution in constraining the corresponding Higgs boson couplings would be negligible in a combined fit.

The contribution of \HH events in the signal regions targeting \singleH events is also studied. This can be nonnegligible for the \ttH-enriched categories because the \ttH cross section is about 16 times larger than the \HH cross section, and because the \HH events with one Higgs boson decaying to a $\bbbar$ quark pair have a final state similar to the \ttH events. The categories targeting \ttH events in the \Hgg and multilepton final states are the most sensitive to this process among the \ttH categories considered for this combination. The \HH contamination in the \ttH-enriched categories is either modeled, as in the \ttHgg categories, or its impact on the results is studied and found to be small. In particular, a possible \HH contamination in the \ttHmultilep analysis categories is estimated to be at most $5\% $ with respect to the total \ttH signal yield in the categories. This possible contamination is expected to cause a shift of the allowed $2\sigma$ \kl interval for the combination of \singleH channels and for the combination of \singleH and \HH channels of at most $2\%$ and $1\%$ with respect to the width of the interval towards higher values of \kl, respectively.

\begin{table*}[ht!]
\centering
	\topcaption{Analyses targeting \singleH production modes and decay channels included in the combination and the corresponding data set sizes, in terms of integrated luminosity. The production modes targeted with dedicated analysis categories and the maximum phase space granularity of each cross section measurement are also reported.}
	\label{tab:singleHanalyses}
	\begin{tabular}{ l  c  c  c  c }
\hline
\multirow{2}{*}{Analysis} & Integrated          & Targeted \PH            & Maximum     & \multirow{2}{*}{References} \\
                          & luminosity (\fbinv) & production modes       & granularity & \\
\hline
\HZZllll & 138 & \ggF, \VBF, \VH, \ttH & STXS~1.2 &~\cite{Sirunyan:2753947} \\
\Hgg & 138 & \ggF, \VBF, \VH, \ttH, \tH & STXS~1.2 &~\cite{Sirunyan:2754281, Sirunyan:2745738} \\
\HWW & 138 & \ggF, \VBF, \VH & STXS~1.2 &~\cite{arXiv:2206.09466} \\
\ttHmultilep & 138 & \ttH & Inclusive &~\cite{Sirunyan:2743741} \\
\ggHbb & 138 & \ggF & Inclusive &~\cite{Sirunyan:2721858} \\
\VHbb & 77 & \VH & Inclusive &~\cite{PhysRevLett.121.121801} \\
\ttHbb & 36 & \ttH & Inclusive &~\cite{Sirunyan:2312382} \\
\Htt & 138 & \ggF, \VBF, \VH & STXS~1.2 &~\cite{arXiv:2204.12957} \\
\Hmm & 138 & \ggF, \VBF & Inclusive &~\cite{Sirunyan:2730058} \\
\hline
    \end{tabular}
\end{table*}

\begin{table*}[ht!]
\centering
	\topcaption{The \HH searches included in the combination and the corresponding data set sizes, in terms of integrated luminosity. The \HH production modes targeted with dedicated event categories are also reported.}
	\label{tab:HHanalyses}
	\begin{tabular}{ l  c  c  c }
\hline
Analysis & Int. luminosity (\fbinv)  & Targeted \HH production modes & References \\
\hline
\HHbbgg & 138 & \ggF and \VBF &~\cite{Sirunyan:2745738} \\
\HHbbtt & 138 & \ggF and \VBF &~\cite{arXiv:2206.09401} \\
\HHbbbb & 138 & \ggF, \VBF, and \VHH &~\cite{Tumasyan:2801882,arXiv:2205.06667, Hayrapetyan2024} \\
\HHmultilep & 138 & \ggF &~\cite{arXiv:2206.10268} \\
\HHbbWW & 138 & \ggF and \VBF &~\cite{CMS:2024rgy} \\
\hline
    \end{tabular}
\end{table*}

\section{Changes to the analyses in the combination}
\label{sec:changes_to_analyses}

One of the main challenges of this combination is estimating the overlaps between the signal regions considered in different analyses and optimizing a strategy for the removal of the nonnegligible overlaps. This is crucial to avoid significant double-counting, ensuring accurate constraints on the Higgs boson trilinear self-coupling and its couplings to fermions and vector bosons. The overlap of \singleH analyses targeting similar final states has been checked and found to be negligible~\cite{arXiv:2207.00043}, while analyses targeting different final states are designed to be orthogonal to each other. The study of the overlap between the different \HH analysis signal regions was performed for the \HH combination presented in Ref.~\cite{arXiv:2207.00043} and the only relevant overlap was found between the searches for \HHbbbb in the resolved-jet and merged-jet final states. An additional veto on events with large-radius jets with $\PT>300$~GeV is applied in the resolved-jet analysis to remove the overlap with minimal impact on the final results. The study of the overlap between \HH analysis regions has been extended to the additional \HH channels introduced in this combination. An additional overlap is found between categories of the \VHHbb analysis targeting fully hadronic \PV decays and signal regions of the \ggHHbbbb analysis. Since the ggF analysis sets stronger constraints on \kl, the hadronic-\PV categories of the \VHH analysis are removed.

Overlaps between \singleH and \HH analyses are identified by comparing the event selection criteria of the analysis subcategories. Two analysis regions are considered to be overlapping if both categories in \singleH and \HH analyses select more than 1\% of the same events in simulated \singleH or \HH event samples. The overlap of \HH events in \singleH analyses is negligible because of the low yield of the \HH signal in the \singleH analysis categories, while significant overlaps are identified for \singleH events reconstructed in the \HH regions. In case two analysis regions are considered overlapping, one of the two is dropped. The removal of the overlap is optimized to provide the highest sensitivity to \kl, without significantly compromising the sensitivity to the Higgs boson couplings to fermions and vector bosons. The \HHbbZZllll analysis~\cite{CMS:2022omp} has not been included in this combination because of its comparatively low sensitivity and high expected overlap with the \HZZllll analysis.

Significant overlaps are generally found between the categories targeting \ttH events and the ones targeting \HH events. In the \Hgg final state, the \ttH STXS~1.2 categories defined in Ref.~\cite{Sirunyan:2754281} are replaced by the inclusive \ttH categories defined in Ref.~\cite{Sirunyan:2745738} for the combined measurement of the \ttHgg and \HHbbgg cross sections, which are by construction orthogonal to each other. In the \ttHmultilep analysis, categories with subleading sensitivity to the \ttH signal are removed because they target final states with the same lepton multiplicity as the signal regions defined in the \HHbbWW or \HHbbtt analyses. In the \ttHbb analysis, the signal regions targeting fully-hadronic decays of the top quark-antiquark pair are removed to prevent overlaps with the \HHbbbb analysis categories.

Overlaps are also found between \HHbbgg analysis categories and some \Hgg analysis categories enriched in \ggH, \qqH, or \VH events. The overlapping \Hgg analysis categories target \ggH or \qqH events with large \PTH, or \VH events with hadronic decays of the vector boson. Those categories are removed. The \HWW analysis categories targeting \VH events in final states with either two same-sign leptons or three leptons are expected to have a nonnegligible overlap with the \HHmultilep analysis categories targeting the same lepton multiplicities. The three-lepton categories of the \HWW analysis and the same-sign dilepton category of the \HHmultilep analysis are dropped. The impact on the coupling constraints of interest is negligible because those categories play a subdominant role in their respective analyses. Overlaps are also found between most of the \Htt analysis categories and the \HHbbtt ones which target \HH events with a \Hbb candidate reconstructed as a pair of jets with exactly one of them identified as a b jet. The overlapping \HHbbtt categories are removed to preserve the sensitivity to the \PH coupling to the tau lepton with minimal impact on the constraints on \kl. Furthermore, the \HHbbtt analysis categories enriched in \singleH events are removed, along with a few remaining \Htt categories targeting \ggH and \qqH events with large \PTH or with two additional jets in the final state.

Finally, overlaps in control regions, which are used to constrain background processes, are studied by comparing selected data events. In cases of significant overlap, one of the overlapping control regions is removed. The removal strategy is optimized to preserve the constraints from data on the background processes and on the related systematic uncertainties. Control regions of the \VHbb and \HHbbWW analyses enriched in \ttbar events are removed because of overlaps with control regions of the \VHHbb and \HWW analyses, respectively, which target similar phase space. The control regions of the \HHmultilep analysis enriched in $\PZ\PW$ and $\PZ\PZ$ events are also removed because of expected overlaps with the signal regions of the \HWW analysis enriched in \VH events.

\section{Statistical procedure}
\label{sec:statistical_methods}

The statistical interpretation of the data is based on the methodology common to the ATLAS and CMS Collaborations~\cite{CMS-PAPERS-HIG-15-002,CMS-NOTE-2011-005} and the results have been obtained using the CMS statistical analysis tool \textsc{Combine}~\cite{CAT-23-001}. A profile likelihood ratio $q$ is used as a test statistic to estimate the parameters of interest (POIs) $\vec{\alpha}$ of the model and their associated confidence intervals, taking into account the systematic uncertainties (Section~\ref{sec:systematic_uncertainties}) as independent nuisance parameters $\vec{\theta}$:
\begin{equation}
	q(\vec{\alpha}) = -2 \log \Biggl( \frac
		{L(\vec{\alpha},       \hat{\hat{\vec{\theta}}}(\vec{\alpha}) )}
		{L(\hat{\vec{\alpha}}, \hat{\vec{\theta}}                )}
	\Biggr),
\end{equation}
where $L$ is the likelihood function, which is built as the product of the individual Poisson probability functions, accounting for the observed data and expected number of events across all the analysis channels, and additional terms representing constraints on the nuisance parameters from external measurements. The POIs and nuisance parameter values maximizing the likelihood function are given by  $(\hat{\vec{\alpha}}, \hat{\vec{\theta}})$, while $ \hat{\hat{\vec{\theta}}}(\vec{\alpha}) $ is the set of nuisance parameter values that maximize the likelihood function for a given set of POI values. Each source of systematic uncertainty corresponds to a nuisance parameter in the likelihood function, possibly correlated among all the input channels affected by that uncertainty.

The $\hat{\vec{\alpha}}$ are taken as the best-fit values of the considered POIs. The intervals at $68.3\% $ and $95.4\% $ \CL correspond to one ($1\sigma$) and two ($2\sigma$) standard deviations, respectively. For one-dimensional measurements, the $1\sigma$ and $2\sigma$ intervals are identified as the regions of POI values for which $q<1$ and $q<4$, respectively, while for two-dimensional measurements they correspond to the regions where $q<2.30$ and $q<6.18$, respectively.

In addition to \kl, this combination takes into consideration the modifiers of the \PH couplings to fermions or vector bosons. The modifiers of the \PH couplings to a pair of fermions or vector bosons, which are referred to as \kF and \kV, respectively, have been introduced in the $\kappa$-framework~\cite{kframework} to parametrize possible deviations of the Higgs boson production cross sections and decay branching fractions from the SM predictions. The \kF and \kV parameters are defined in such a way that the deviations from the SM prediction
for a \SH production mechanism $i$ and its cross section $\sigma_i$, or a  \PH decay channel $j$ and the corresponding partial decay width $\Gamma_j$, is parametrized as:
\begin{equation}\label{eq:H_coupling_modifiers}
\frac{\sigma_i}{\sigma_i^{\text{SM}}} = \kappa_i^2 \text{ or } \frac{\Gamma_j}{\Gamma_j^{\text{SM}}} = \kappa_j^2.
\end{equation}
The loop-induced \PH production via gluon fusion and \PH decay to a pair of photons or gluons are parametrized as a function of the \PH couplings to the fermions or vector bosons entering in the loop. This combination alternatively considers the \kF parameter, which coherently modifies the \PH couplings to all fermions, or the \kappat, \kb, \kmu, and \ktau parameters which independently modify the couplings to the corresponding fermions. In this second case, the \PH couplings to the remaining fermions are assumed to be fixed to their SM predictions because no analyses sensitive to such parameters are included in the combination.

The modifier of the SM couplings between two Higgs bosons and two massive vector bosons, $\kVV$,  is used to describe possible deviations from the SM expectation for the \HH production from VBF mechanism. The cross sections of the \singleH processes considered in this combination do not depend on \kVV at the leading order, while effects at higher perturbative orders are assumed negligible. One or at most two coupling modifiers among \kl, \kappat, \kVV, and \kV are chosen as the POIs in this combination. The remaining coupling modifiers are assumed to be either equal to the SM prediction or treated as unconstrained nuisance parameters.

The modeling of the dependence of the single-H cross sections and branching fractions on \kl are provided in Refs~\cite{Degrassi:2016wml,arXiv:1709.08649v2,LHCHWG-2022-002}. Those parametrizations are valid under the assumption of small deviations of \kV and \kF from unity. The reason is that the parametrizations are based on an incomplete perturbative order expansion, which is at LO in \kV and \kF and at NLO in electroweak theory in \kl. This is a good approximation for deviations of \kV and \kF within a few percent from unity because in this range the missing higher-order terms would have a negligible effect on the cross sections and branching fractions compared to the experimental and theoretical uncertainties.

\section{Systematic uncertainties}\label{sec:systematic_uncertainties}

The systematic uncertainties, which affect the modeling of a signal or a background process, are described as uncertainties in the shape of the distributions or normalization uncertainties in the yield of the affected processes. The systematic uncertainties may originate either from the theoretical predictions or from the measurement. The most important uncertainties are described in the next paragraphs. The \HH production searches are limited by the statistical power of the data.  In the combination of \HH channels, the overall impact of the systematic uncertainties on the constraints on \kl is about $8\% $ of the total uncertainty. The most impactful systematic uncertainties are in the theory predictions of the \ggHH cross section, and in the estimations from data of important background components in the \bbbb and \bbtt channels. In the \singleH channels, the systematic uncertainties tend to be more important. The overall impact of the systematic uncertainties on the constraints on \kl from the combination of \singleH channels is about $30\% $.  The leading uncertainties are in the theory predictions of the \ggH and \ttH production cross sections \cite{cernYR4}. Several modifications to the input channels are performed to use a consistent uncertainty model similar to the uncertainty model in the \singleH combination presented in Ref.~\cite{arXiv:2207.00043}. The impact of all these modifications on the results is found to be negligible.

The uncertainties in the theory predictions of the \singleH and \HH cross sections are among the dominant sources of systematic uncertainty on \kl. The uncertainties in the \HH and \singleH cross sections originate from the renormalization and factorization scales, the parton distribution functions (PDF), as well as the branching fraction predictions. An additional important source of uncertainty in the \ggHH cross section is related to the renormalization scheme and scale dependence of the virtual top quark mass~\cite{PhysRevD.103.056002}. Such theory uncertainties are fully correlated among all affected channels as are the uncertainties in the theoretical prediction of the cross section of a background process, whenever a similar phase space is considered. In the \singleH channels the uncertainty scheme for the \ggH cross section is defined according to Ref.~\cite{cernYR4}. This consists of a set of nuisance parameters that cover for the uncertainty in the total \ggH cross section, as well as the cross sections in the exclusive phase space regions targeted by the STXS framework. The \HH analyses with a nonnegligible contamination  from \ggH production do not perform any explicit classification of the \ggH events in STXS bins. Therefore, in the \HH channels only the uncertainty in the total \ggH cross section is taken into account and correlated with the \singleH channels.

Uncertainties affecting multiple channels such as those in the integrated luminosity, in the reconstruction and identification of photons, electrons, muons, hadronically decaying tau leptons, in the b jet identification, and in the jet energy scale and resolution are correlated among multiple channels, following the same procedure as for the combinations presented in Ref.~\cite{arXiv:2207.00043}. Additional experimental uncertainties related to measurement procedures specific to each channel, which are detailed in the dedicated publications (listed in Tables~\ref{tab:singleHanalyses} and \ref{tab:HHanalyses}), are uncorrelated among the channels.

\section{Results}
\label{sec:results}

\begin{figure}[ht!]
	\centering
	\includegraphics[width=.45\textwidth]{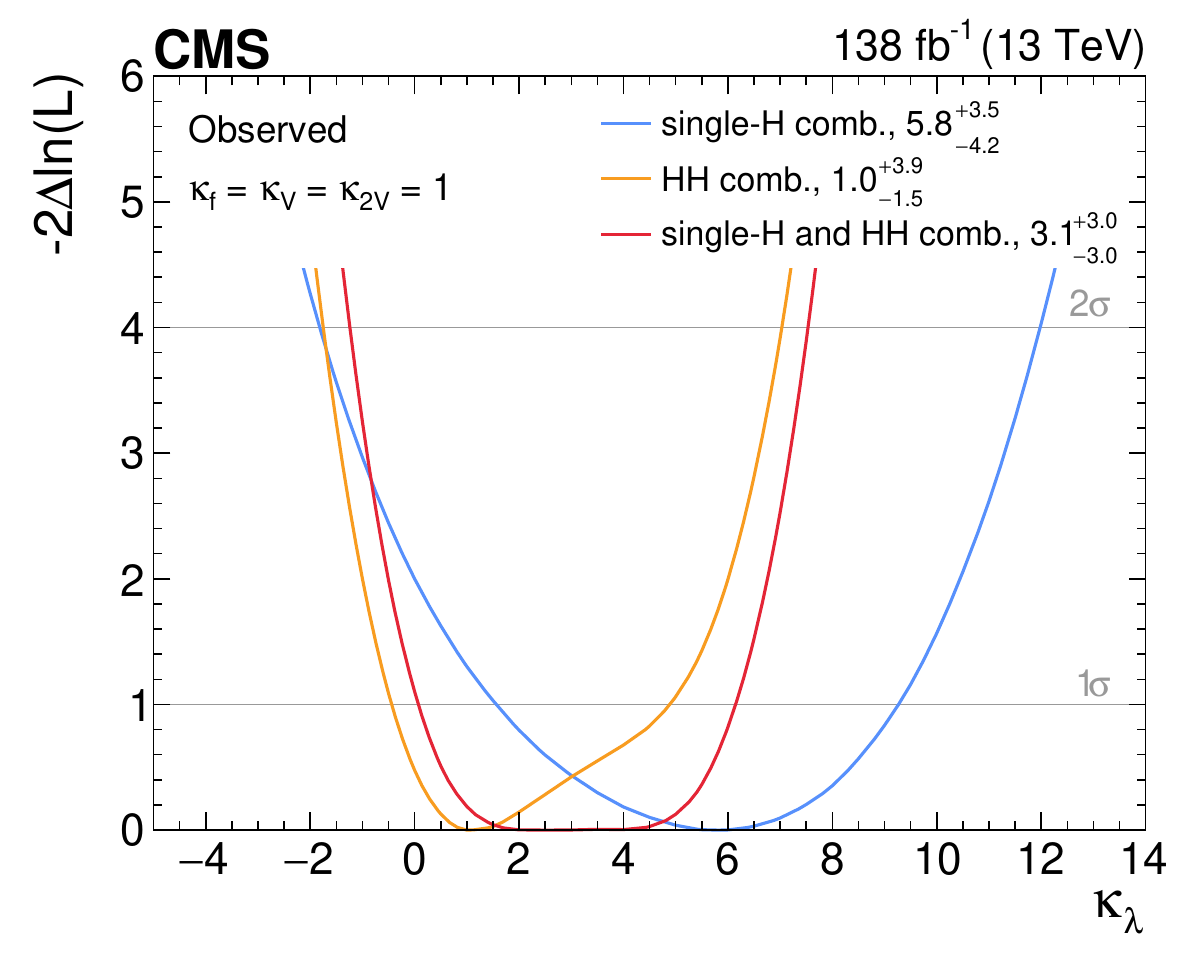}
	\caption{Observed profile likelihood scans of \kl comparing the full combination of \singleH and \HH (red) to the combinations of only \singleH (blue) or only \HH (yellow)  channels. The legend includes the best-fit values and the $1\sigma$ uncertainties.}
	\label{fig:klscan_fixingPOIs}
\end{figure}

The profile likelihood scan of the \kl parameter is shown in Fig.~\ref{fig:klscan_fixingPOIs}, comparing the separate combinations of the \singleH or \HH channels to the full combination of the two, assuming in all cases the other \PH couplings equal to their SM predictions. The observed best-fit \kl from the combination of the \HH channels is found very close to unity, and the observed $2\sigma$ allowed interval is found to be $-1.7<\kl<7.0$, in agreement with the expected interval of $-2.3<\kl<8.0$, assuming the other Higgs boson couplings fixed to the SM prediction. The observed best-fit \kl from the combination of the \SH channels is larger than one, although compatible with the SM within the $2\sigma$ uncertainty. The observed and expected $2\sigma$ allowed intervals are $-1.8<\kl<12.0$ and $-4.5<\kl<11.7$, respectively. The profile likelihood obtained considering both the \SH and \HH channels is qualitatively flat around the minimum. The shift in the \kl interval from the \SH combination is mostly driven by the \ttHgg channel. The observed (expected) $2\sigma$ allowed interval from \SH and \HH combination is  $-1.2<\kl<7.5$ ($-2.0<\kl<7.7$). The observed allowed interval on \kl can be compared to the corresponding one from the \HH combination of Ref.~\cite{arXiv:2207.00043}. The lower bounds are compatible, while the upper bound from this combination is weaker because of the \SH channels shifting the interval towards higher values of \kl. The observed allowed interval on \kl from the combination of only the \HH channels can also be compared to Ref.~\cite{arXiv:2207.00043}. The differences, including the change in the observed best-fit \kl value, which moves closer to the SM prediction, are the result of the changes described in Section~\ref{sec:changes_to_analyses} and most notably the removal of categories from the \HHbbtt in favor of the \Htt analysis. It should be noted that any loss of sensitivity in the \HH combination as a result of those changes is fully recovered in the combination with the \SH channels.

\begin{figure}[ht!]
	\centering
	\includegraphics[width=.45\textwidth]{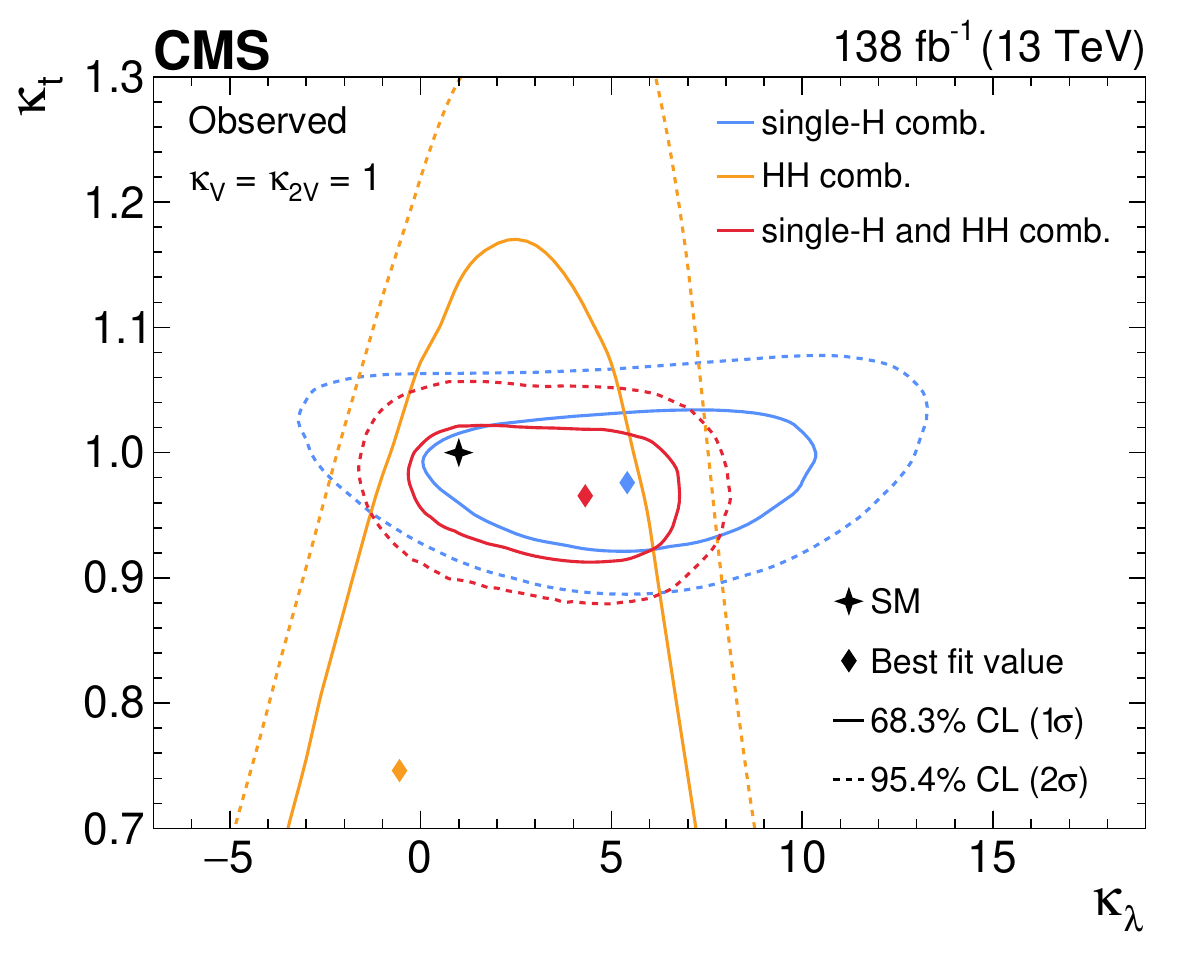}
	\caption{Observed two-dimensional likelihood scans of $(\kl, \kappat)$  comparing the full combination of \singleH and \HH (red) to the combinations of only \singleH (blue) or only \HH (yellow) channels.}
	\label{fig:klktscan}
\end{figure}

The two-dimensional likelihood scan of the \kl and \kappat parameters is shown in Fig.~\ref{fig:klktscan}. A large degeneracy of the \ggHH cross section with respect to \kl and \kappat limits the \kl sensitivity of the \HH channels. In that case,  the constraints on \kappat arise mostly from the contamination of \singleH events in the \HH enriched categories, and from the \kappat dependence of the \PH branching fractions, especially in the \Hgg channel. Instead, the \singleH combination provides a stringent constraint on \kappat, which is utilized in the combination of the \singleH and \HH channels.

\begin{figure}[ht!]
	\centering
	\includegraphics[width=.4\textwidth]{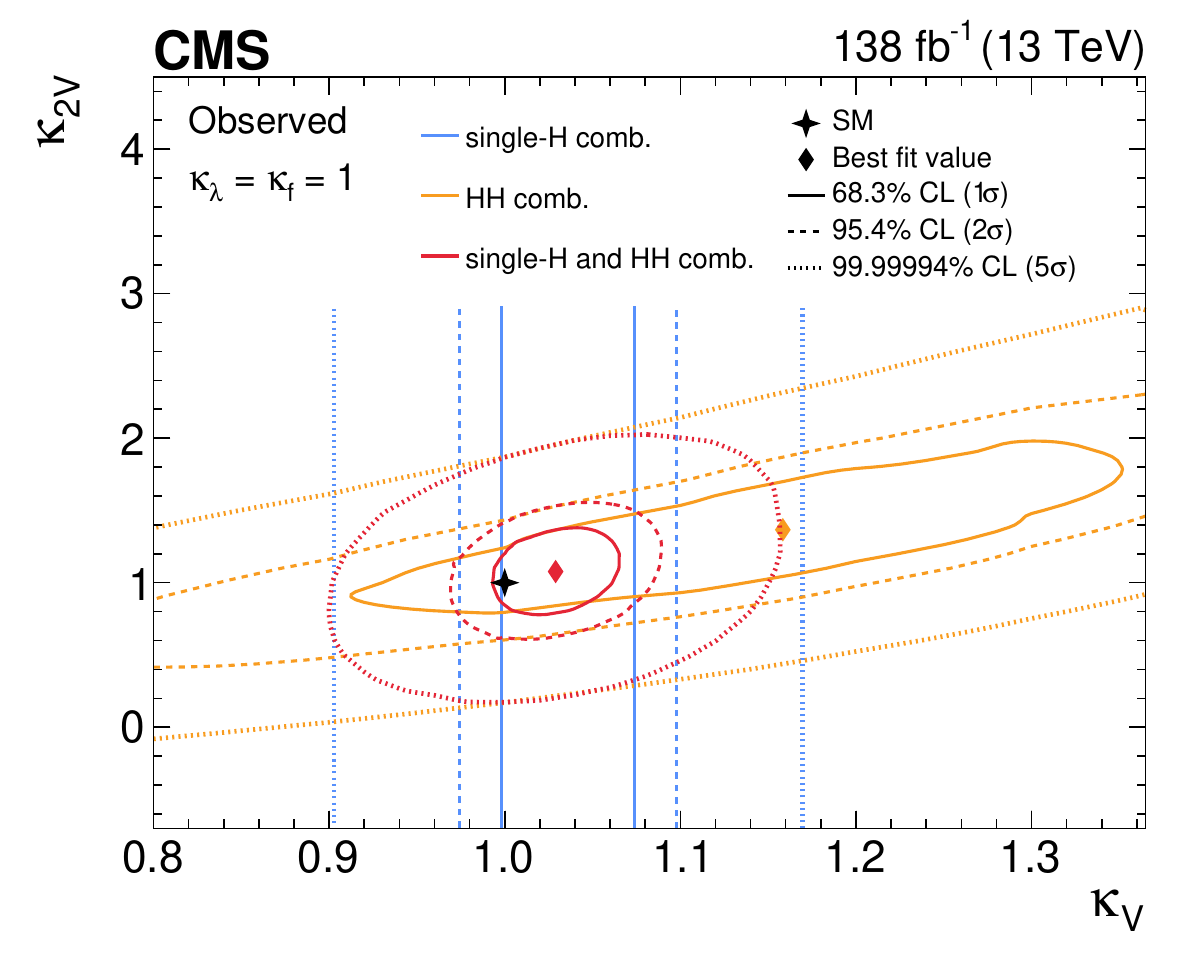}
	\caption{Observed two-dimensional likelihood scans of $(\kV, \ktwoV)$ comparing the full combination of \singleH and \HH (red) to the combinations of only \singleH (blue) or only \HH (yellow) channels. The \singleH combination has no sensitivity on the \ktwoV parameter.}
	\label{fig:kVk2Vscan}
 \end{figure}

The complementarity of the \singleH and \HH channels is also employed in the simultaneous measurement of the \kV and \ktwoV parameters. The two-dimensional likelihood scan of the \kV and \ktwoV parameters is shown in Fig.~\ref{fig:kVk2Vscan}. The constraint on \ktwoV is driven by the \HH categories enriched in \qqHH events. However, the \qqHH cross section has a large degeneracy with respect to the \kV and \ktwoV parameters. The \singleH channels have no sensitivity on the \ktwoV parameter but provide a stringent constraint on \kV. This allows the exclusion of $\ktwoV \leq 0$ for any value of \kV with a significance larger than five standard deviations, under the assumption that \kVV contributions to the single H cross section are negligible in the considered range. 

\begin{figure}[ht!]
	\centering
	\includegraphics[width=.4\textwidth]{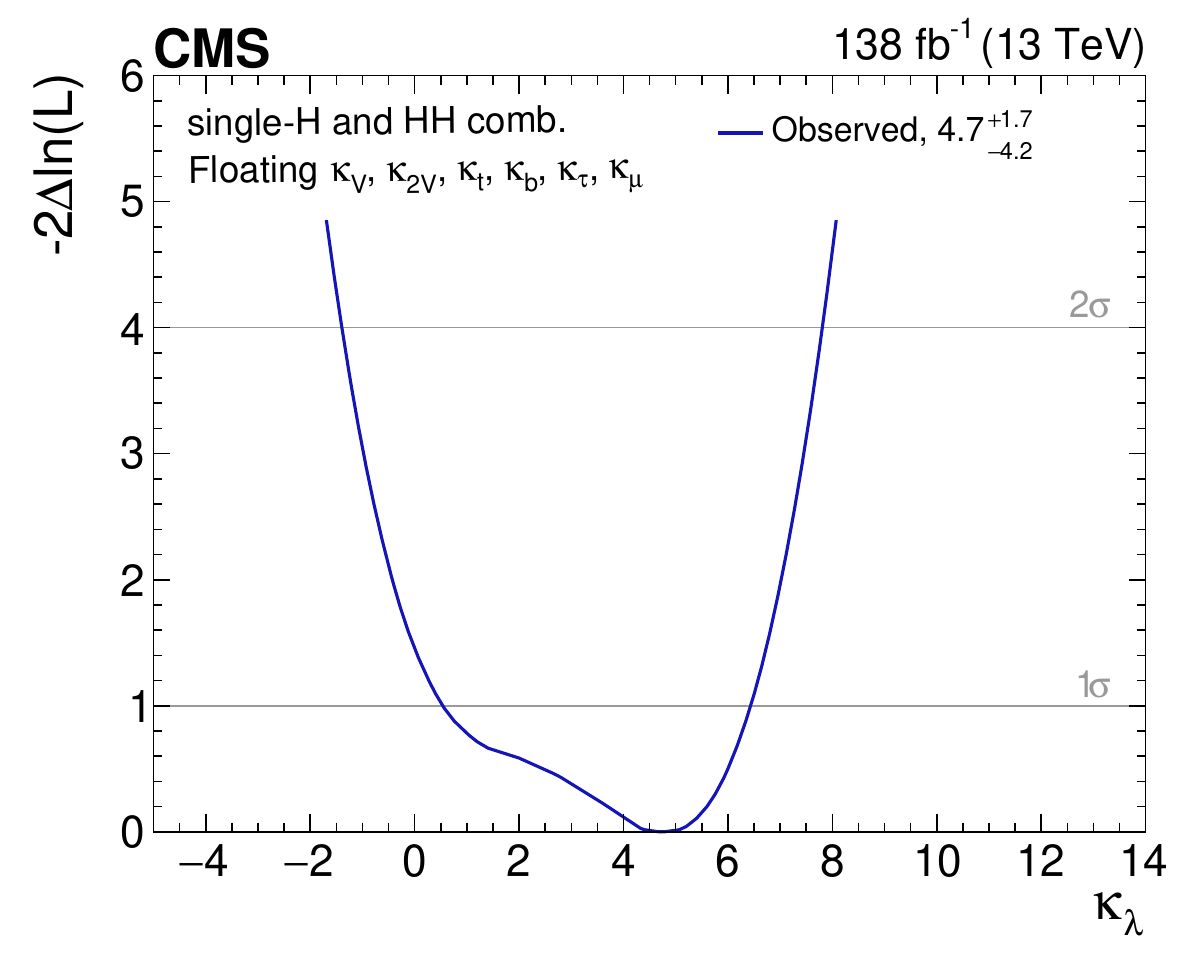}
	\caption{Observed likelihood scans of \kl assuming \kV, \ktwoV, \kappat, \kb, \ktau, and \kmu as unconstrained nuisance parameters. The legend includes the best-fit value and the $1\sigma$ uncertainty.}
	\label{fig:klscan_floatingPOIs}
\end{figure}

Constraints on \kl are also set under less restrictive assumptions on the Higgs boson couplings to the fermions and vector bosons. In particular, the \kl likelihood scan treating \kV, \ktwoV, \kappat, \kb, \ktau, and \kmu as unconstrained nuisance parameters is shown in Fig.~\ref{fig:klscan_floatingPOIs}. The degeneracy that would be present if only \HH categories were considered is fully resolved in the combination of the \singleH and \HH channels. The resulting expected constraint on \kl is comparable to the one expected from the \HH combination assuming the other Higgs boson couplings equal to their SM value. The constraints on \kl under the aforementioned assumptions, along with some alternative assumptions, including the ones used by the ATLAS Collaboration in Ref.~\cite{ATLAS:2022jtk}, are summarized in Table~\ref{tab:klconstraints}. The results are found to be consistent with the combination of \singleH and \HH measurements and searches from the ATLAS Collaboration~\cite{ATLAS:2022jtk}.

\begin{table*}[htb!]
\centering
	\topcaption{Expected and observed constraints on \kl at $2\sigma$ and best fit values from the combination of the \singleH and \HH channels under different assumptions on the Higgs boson couplings to fermions and vector bosons. The floating coupling parameters are either treated as POIs or as unconstrained nuisance parameters in the fit. The parameters that are not listed as floating are always assumed fixed to the SM prediction.}
	\label{tab:klconstraints}
	\begin{tabular}{ l  r  r  r  r }	
\hline
  & \multicolumn{2}{c}{Best fit \kl value $\pm 1\sigma$} & \multicolumn{2}{c}{$2\sigma$ interval} \\ 
Hypothesis & Expected & Observed & Expected & Observed \\
\hline
Other couplings fixed to the SM prediction & $1.0^{+4.6}_{-1.7}$ & $3.1^{+3.0}_{-3.0}$ & $[-2.0, 7.7]$ & $[-1.2, 7.5]$ \\
Floating (\kV, \ktwoV, \kF) & $1.0^{+4.7}_{-1.8}$ & $4.5^{+1.8}_{-4.7}$ & $[-2.2, 7.8]$ & $[-1.7, 7.7]$ \\
Floating (\kV, \kappat, \kb, \ktau) & $1.0^{+4.8}_{-1.8}$ & $4.7^{+1.7}_{-4.1}$ & $[-2.3, 7.7]$ & $[-1.4, 7.8]$ \\
Floating (\kV, \ktwoV, \kappat, \kb, \ktau, \kmu) & $1.0^{+4.8}_{-1.8}$ & $4.7^{+1.7}_{-4.2}$ & $[-2.3, 7.8]$ & $[-1.4, 7.8]$ \\
\hline
    \end{tabular}
\end{table*}

\section{Summary}
\label{sec:summary}

A combination of Higgs boson (\PH) measurements and searches for Higgs boson pair production (\HH) to constrain the Higgs boson trilinear self-coupling has been presented. Proton-proton collision data at $\sqrt{s} = 13\TeV$, collected by the CMS experiment between 2016 and 2018, were analyzed. This is the first combination of \singleH and \HH channels from the CMS Collaboration. The complementarity of the constraints on the Higgs boson couplings of the \singleH and \HH channels is employed in the combination. The \singleH channels lessen the assumptions on the measurements on the modifier of the Higgs boson trilinear self-coupling \kl, allowing for simultaneous constraint of the Higgs boson couplings to fermions and vector bosons. The observed (expected) interval at $2\sigma$ confidence level on \kl under the assumption that the other Higgs boson couplings are fixed to the standard model prediction, is found to be $-1.2<\kl<7.5$ ($-2.0<\kl<7.7$). Relaxing the assumption on the Higgs couplings to fermions and vector bosons the corresponding observed (expected) interval on \kl is $-1.4<\kl<7.8$ ($-2.3<\kl<7.8$).

\begin{acknowledgments}	We congratulate our colleagues in the CERN accelerator departments for the excellent performance of the LHC and thank the technical and administrative staffs at CERN and at other CMS institutes for their contributions to the success of the CMS effort. In addition, we gratefully acknowledge the computing centers and personnel of the Worldwide LHC Computing Grid and other centers for delivering so effectively the computing infrastructure essential to our analyses. Finally, we acknowledge the enduring support for the construction and operation of the LHC, the CMS detector, and the supporting computing infrastructure provided by the following funding agencies: SC (Armenia), BMBWF and FWF (Austria); FNRS and FWO (Belgium); CNPq, CAPES, FAPERJ, FAPERGS, and FAPESP (Brazil); MES and BNSF (Bulgaria); CERN; CAS, MoST, and NSFC (China); MINCIENCIAS (Colombia); MSES and CSF (Croatia); RIF (Cyprus); SENESCYT (Ecuador); ERC PRG, RVTT3 and MoER TK202 (Estonia); Academy of Finland, MEC, and HIP (Finland); CEA and CNRS/IN2P3 (France); SRNSF (Georgia); BMBF, DFG, and HGF (Germany); GSRI (Greece); NKFIH (Hungary); DAE and DST (India); IPM (Iran); SFI (Ireland); INFN (Italy); MSIP and NRF (Republic of Korea); MES (Latvia); LMTLT (Lithuania); MOE and UM (Malaysia); BUAP, CINVESTAV, CONACYT, LNS, SEP, and UASLP-FAI (Mexico); MOS (Montenegro); MBIE (New Zealand); PAEC (Pakistan); MES and NSC (Poland); FCT (Portugal);  MESTD (Serbia); MCIN/AEI and PCTI (Spain); MOSTR (Sri Lanka); Swiss Funding Agencies (Switzerland); MST (Taipei); MHESI and NSTDA (Thailand); TUBITAK and TENMAK (Turkey); NASU (Ukraine); STFC (United Kingdom); DOE and NSF (USA).

\hyphenation{Rachada-pisek} Individuals have received support from the Marie-Curie program and the European Research Council and Horizon 2020 Grant, contract Nos.\ 675440, 724704, 752730, 758316, 765710, 824093, 101115353, 101002207, and COST Action CA16108 (European Union); the Leventis Foundation; the Alfred P.\ Sloan Foundation; the Alexander von Humboldt Foundation; the Science Committee, project no. 22rl-037 (Armenia); the Belgian Federal Science Policy Office; the Fonds pour la Formation \`a la Recherche dans l'Industrie et dans l'Agriculture (FRIA-Belgium); the Agentschap voor Innovatie door Wetenschap en Technologie (IWT-Belgium); the F.R.S.-FNRS and FWO (Belgium) under the ``Excellence of Science -- EOS" -- be.h project n.\ 30820817; the Beijing Municipal Science \& Technology Commission, No. Z191100007219010 and Fundamental Research Funds for the Central Universities (China); the Ministry of Education, Youth and Sports (MEYS) of the Czech Republic; the Shota Rustaveli National Science Foundation, grant FR-22-985 (Georgia); the Deutsche Forschungsgemeinschaft (DFG), under Germany's Excellence Strategy -- EXC 2121 ``Quantum Universe" -- 390833306, and under project number 400140256 - GRK2497; the Hellenic Foundation for Research and Innovation (HFRI), Project Number 2288 (Greece); the Hungarian Academy of Sciences, the New National Excellence Program - \'UNKP, the NKFIH research grants K 131991, K 133046, K 138136, K 143460, K 143477, K 146913, K 146914, K 147048, 2020-2.2.1-ED-2021-00181, and TKP2021-NKTA-64 (Hungary); the Council of Science and Industrial Research, India; ICSC -- National Research Center for High Performance Computing, Big Data and Quantum Computing and FAIR -- Future Artificial Intelligence Research, funded by the NextGenerationEU program (Italy); the Latvian Council of Science; the Ministry of Education and Science, project no. 2022/WK/14, and the National Science Center, contracts Opus 2021/41/B/ST2/01369 and 2021/43/B/ST2/01552 (Poland); the Funda\c{c}\~ao para a Ci\^encia e a Tecnologia, grant CEECIND/01334/2018 (Portugal); the National Priorities Research Program by Qatar National Research Fund; MCIN/AEI/10.13039/501100011033, ERDF ``a way of making Europe", and the Programa Estatal de Fomento de la Investigaci{\'o}n Cient{\'i}fica y T{\'e}cnica de Excelencia Mar\'{\i}a de Maeztu, grant MDM-2017-0765 and Programa Severo Ochoa del Principado de Asturias (Spain); the Chulalongkorn Academic into Its 2nd Century Project Advancement Project, and the National Science, Research and Innovation Fund via the Program Management Unit for Human Resources \& Institutional Development, Research and Innovation, grant B37G660013 (Thailand); the Kavli Foundation; the Nvidia Corporation; the SuperMicro Corporation; the Welch Foundation, contract C-1845; and the Weston Havens Foundation (USA).	
\end{acknowledgments}

\bibliography{auto_generated}
\cleardoublepage \appendix\section{The CMS Collaboration \label{app:collab}}\begin{sloppypar}\hyphenpenalty=5000\widowpenalty=500\clubpenalty=5000
\cmsinstitute{Yerevan Physics Institute, Yerevan, Armenia}
{\tolerance=6000
A.~Hayrapetyan, A.~Tumasyan\cmsAuthorMark{1}\cmsorcid{0009-0000-0684-6742}
\par}
\cmsinstitute{Institut f\"{u}r Hochenergiephysik, Vienna, Austria}
{\tolerance=6000
W.~Adam\cmsorcid{0000-0001-9099-4341}, J.W.~Andrejkovic, T.~Bergauer\cmsorcid{0000-0002-5786-0293}, S.~Chatterjee\cmsorcid{0000-0003-2660-0349}, K.~Damanakis\cmsorcid{0000-0001-5389-2872}, M.~Dragicevic\cmsorcid{0000-0003-1967-6783}, P.S.~Hussain\cmsorcid{0000-0002-4825-5278}, M.~Jeitler\cmsAuthorMark{2}\cmsorcid{0000-0002-5141-9560}, N.~Krammer\cmsorcid{0000-0002-0548-0985}, A.~Li\cmsorcid{0000-0002-4547-116X}, D.~Liko\cmsorcid{0000-0002-3380-473X}, I.~Mikulec\cmsorcid{0000-0003-0385-2746}, J.~Schieck\cmsAuthorMark{2}\cmsorcid{0000-0002-1058-8093}, R.~Sch\"{o}fbeck\cmsorcid{0000-0002-2332-8784}, D.~Schwarz\cmsorcid{0000-0002-3821-7331}, M.~Sonawane\cmsorcid{0000-0003-0510-7010}, S.~Templ\cmsorcid{0000-0003-3137-5692}, W.~Waltenberger\cmsorcid{0000-0002-6215-7228}, C.-E.~Wulz\cmsAuthorMark{2}\cmsorcid{0000-0001-9226-5812}
% ...
\par}
\vskip\cmsinstskip
\dag:~Deceased\\
$^{1}$Also at Yerevan State University, Yerevan, Armenia\\
$^{2}$Also at TU Wien, Vienna, Austria\\
% ...
\end{sloppypar}
\end{document}